# Relaxation in the glass-former acetyl salicylic acid studied by deuteron magnetic resonance and dielectric spectroscopy

R. Nath,[1] T. El Goresy,[1] H. Zimmermann,[2] R. Böhmer[1]

[1] *Experimentelle Physik III and Interdisziplinäres Zentrum für Magnetische Resonanz, Universität Dortmund, 44221 Dortmund, Germany*

[2] *Max-Planck-Institut für Medizinische Forschung, 69120 Heidelberg, Germany*

Supercooled liquid and glassy acetyl salicylic acid was studied using dielectric spectroscopy and deuteron relaxometry in a wide temperature range. The supercooled liquid is characterized by major deviations from thermally activated behavior. In the glass the secondary relaxation exhibits the typical features of a Johari-Goldstein process. Via measurements of spin-lattice relaxation times the selectively deuterated methyl group was used as a sensitive probe of its local environments. There is a large difference in the mean activation energy in the glass with respect to that in crystalline acetyl salicylic acid. This can be understood by taking into account the broad energy barrier distribution in the glass.

PACS number: 76.60.–k, 64.70.Pf, 77.22.–d, 87.64.Fb

Many molecular crystals show a rich polymorphism.[1] This is particularly important for substances of pharmaceutical interest, such as caffeine[2] or acetyl salicylic acid (ASA, also known as aspirin),[3] since different crystal forms typically exhibit different physical properties. This can make it difficult to industrially process such substances in a reproducible fashion. One possible solution to the problem is to avoid the handling of crystalline phases altogether and to work with vitreous materials. In some pharmaceutically relevant cases this gives additional benefits such as, e.g., an increased solubility of vitreous medications under physiological conditions.[4] Of course, to be of practical interest the glass transition of the corresponding materials has to be higher than room temperature. Therefore ASA, the substance we investigate in the present work, is not suitable in this respect since its glass transition temperature $T_g$ is at 243 K.[5] Thus while crystalline ASA has been the subject of numerous investigations, the supercooled liquid form of ASA has only rarely been studied.[6,7,8] From a physics point of view the main interest in the crystalline material concerns the dynamics of the methyl group of ASA which has been studied by neutron scattering[9,10] or by nuclear magnetic resonance (NMR) using deuterons[11,12] or carbons[13] as nuclear probes. These studies have shown that the methyl group rotation remains fast on a nanosecond time scale, down to below 50 K.

With the expectation that this motion may be similarly fast in vitreous ASA, we performed $^2$H-NMR on a sample with selectively labeled $CD_3$ groups. As we show, this expectation can be confirmed and enables an observation of the Johari-Goldstein (JG) secondary relaxation[14] via measurements of the deuteron spin-lattice relaxation $T_1$. Such studies on pure small-molecule glass formers are often hampered by the interference of spin-diffusion effects with long $T_1$ times.[15] In vitreous ASA the untypically short $T_1$, shared by only very few other neat organic glasses,[16] arises from the fast methyl group rotation and allows one the study of the glassy dynamics down to low temperature. ASA is particularly attractive in this respect because it exhibits a large separation of $T_g$ (at which the structural relaxation time is about 100 s) and the temperature at which the dynamics of the $CD_3$ group crosses the NMR time scale. The latter is a few nanoseconds and is defined by the inverse of the Larmor frequency $\omega_L$ which in the present work is 46.2 MHz. Thus, ASA appears as one of the few model systems which allows one to compare the crystalline with the local vitreous dynamics.

For most of our studies we used ASA-$d_3$ (HCOO-C$_6$H$_4$-COOCD$_3$) freshly synthesized as described previously.[17] The degree of deuteration of the methyl group was more than 97%. For the dielectric investigations also fully protonated ASA, from Sigma-Aldrich Co. (purity 99.5%), was used. The NMR measurements were carried out using a home-built spectrometer. In addition, the primary and secondary relaxation of ASA was characterized by dielectric and stimulated-echo NMR spectroscopy, in order to check whether it exhibits typical features of small-molecule glass-forming liquids. For the measurement of the complex dielectric constant, $\varepsilon^* = \varepsilon' - i\varepsilon''$, equipment from Novocontrol Co. was used, covering a broad frequency range, $10^{-1}$ Hz $\leq \nu \leq 10^6$ Hz.

Near room temperature ASA exhibits a strong tendency to crystallize.[6] For all NMR measurements ASA-$d_3$ was sealed in a glass tube and slowly heated above its melting point $T_m$ (~407-409 K). For the experiments carried out above ambient temperature no further procedures were required, for those below ambient the molten samples were quenched to ice water. For the dielectric measurements the ASA samples were carefully heated on brass electrodes to just above $T_m$, then squeezed between this and another preheated electrode, and cooled down to 240 K in about 10 min. From there the sample temperature was ramped with rates of ±1 K/min to 115 K or to 290 K.

Typical dielectric loss factor spectra $\tan\delta(\omega) = \varepsilon''(\omega)/\varepsilon'(\omega)$ of ASA-$d_3$ as obtained for below $T_g$ are shown in Fig. 1. It is seen how the symmetrically broadened peaks shift through the experimental frequency window as the temperature is changed. The data could be described using the Cole-Cole function,[18] $\varepsilon^*(\nu) = \varepsilon_\infty + \Delta\varepsilon\, 2\pi\nu\tau_\beta\, [1 + (2\pi i\nu\tau_\beta)^\alpha]$. Here, $\Delta\varepsilon$ is the relaxation strength and $\varepsilon_\infty$ the high-frequency permittivity. $\tau_\beta$ denotes the (mean) correlation time corresponding to the JG-$\beta$-process and $\alpha$ is a parameter describing its distribution. The latter changes from 0.25 to 0.16 when going from T = 230 K to 160 K. This corresponds to a relatively large spectral width which is 6 decades at T = 230 K and roughly follows a 1/T dependence, typical for a distribution of energy barriers with a fixed width.

The correlation times $\tau_\beta$ follow an Arrhenius law, $\tau_\beta(T) = \tau_{\infty,\beta}\exp(E_\beta/T)$, see Fig. 2. From measurements on protonated and on deuterated samples the prefactor turns out to be $\tau_{\infty,\beta} = 3\times10^{-15}$ s and the energy barrier (in temperature units) is $E_\beta = (4900 \pm 50)$ K. Thus, we find $E_\beta/T_g \approx 20.4$ which is compatible with the ratio expected for what has been called a genuine JG process.[19] In Fig. 2 we summarize also the correlation times that were obtained in the range of the primary relaxation. Here differential scanning calorimetry[5,6] and dielectric measurements from this and from previous[6] work are used. Also data from stimulated-echo spectroscopy[20] and from spin-lattice relaxometry are included. Within experimental error the temperature dependences of these time constants agree with one another[21] and can be described by the Vogel-Fulcher expression, $\tau_\alpha = \tau_{\infty,\alpha}\exp[B/(T - T_0)]$, cf. the solid line in Fig. 2. Here the parameters are $B = 2035 \pm 50$ K, $T_0 = 190 \pm 3$ K and $\log_{10}(\tau_{\infty,\alpha}/s) = -15 \pm 0.3$. Using these parameters a steepness index[22] $m = 79 \pm 5$ was determined and $\tau_{\infty,\alpha} = 100$ s obtained at $242 \pm 3$ K is in good agreement with the calorimetric $T_g$.

In Fig. 3 we present the results from our spin-lattice relaxation measurements. $T_1$ was obtained from a fit of the longitudinal magnetization recovery, $M(t)$, detected subsequent to inversion or saturation. For all temperatures shown in this work $M(t)$ could be described well using a Kohlrausch function, $M(t) \propto \exp[-(t/T_1)^{1-\nu}]$. The mean spin-lattice relaxation time can be calculated via $\langle T_1 \rangle = (1-\nu)^{-1} T_1 \Gamma[(1-\nu)^{-1}]$ where $\nu$ is a stretching exponent and $\Gamma$ denotes Euler's gamma function. The exponent $\nu$ is a measure for the width of the distribution $g(T_1)$ of spin-lattice relaxation times with $\int g(T_1)dT_1 = 1$. When cooling down from high temperatures $\langle T_1 \rangle$ passes through a minimum when the correlation time fulfills the condition $\tau_\alpha \sim 0.6/\omega_L \approx 2$ ns. This information is represented as an open diamond in Fig. 2. The NMR correlation time is fully compatible with that extrapolated from the dielectric measurements carried out at lower temperatures. This indicates that the slow-down of the overall molecular tumbling seen in NMR is related to the $\alpha$-process. The value of $T_1$ in the minimum is relatively long and indicative for a methyl group motion which is much faster than $\tau_\alpha$.[16]

For $T > T_g$ any dynamic heterogeneities are successively averaged out and the exponent $\nu$ vanishes within experimental error. For lower temperatures $\nu$ turns non-zero indicating the ergodicity breaking associated with vitrification.[23] The unusual feature, not reported so far for other pure small-molecule glass formers, is that the stretching parameter $\nu$ (Fig. 3b) steadily increases over a very wide temperature range for $T < T_g$.

The spin-lattice relaxation of glassy ASA-$d_3$ varies in a thermally activated fashion and this is also true for crystalline counterpart. The $T_1$ measurements of the crystals, Fig. 3, were conducted at different Larmor frequencies: Nevertheless they almost coincide since in the temperature range shown in Fig. 3 the anisotropic motion of the methyl group is fast with respect to $\omega_L$. The minor difference between the data of Ref. 11 and 12 is due to the fact that $T_1$ depends, albeit only slightly, on the crystal orientation with respect to the external magnetic field. Consequently, the magnetization recovery is very close to exponential. The hindering barrier characterizing the threefold methyl group reorientation in the crystal is 510 K.[11] The effective barrier obtained for glassy ASA from the temperature dependence of $\langle T_1 \rangle$ is much lower, see Fig. 3, but nevertheless $\langle T_1 \rangle$ is longer. Obviously the dynamics of the $CD_3$ group sensitively depends on the local environment and thus can successfully be used to probe the glass structure.

The question arises whether it is the existence of some highly distorted $CD_3$ potentials that leads to the relatively long $T_1$ components in the glass. These would then dominate the time average $\langle T_1 \rangle = \int g(T_1) T_1 dT_1 = \int M(t) dt$ to which spin-lattice relaxometry is sensitive. The short-time end of this distribution is roughly given by the rate average $\langle 1/T_1 \rangle = \int dT_1 \, g(T_1)/T_1$.[24] However, due to the diverging initial slope of the stretched exponential function, the rate average is not defined for the corresponding distribution function. But the assumption of a log-normal form of $g(T_1)$ with a width $\Delta$ (in decades) yields $\langle T_1 \rangle \langle 1/T_1 \rangle = \exp[(\Delta/1.02)^2]$.[24] Now taking a stretching of $\nu = 0.5$, which is obtained near 100 K, the associated width[25] $\Delta$ is about two decades and $\langle T_1 \rangle \langle 1/T_1 \rangle$ is 45. From Fig. 3 one can estimate a ratio of $\langle T_1 \rangle$ for the glass to the rate $1/T_1 = \langle 1/T_1 \rangle$ for the crystal of about 10 near this temperature. This rough agreement indicates that the mean spin-lattice relaxation rates, and thus the smallest barriers against methyl group rotation, are similar in the glass and in the crystal.

The use of a log-normal distribution of time constants, corresponding to a Gaussian distribution of energy barriers, is not unreasonable in the present context. It predicts a continuous broadening of the width of the relaxation time spectrum, in harmony with what is obvious from Fig. 3. The relatively broad distribution reflects the dynamic heterogeneity resulting from the frozen in disorder characterizing the glass.

In summary, we successfully studied the relaxation dynamics in the amorphous form of ASA above and below $T_g$. For $T > T_g$ we obtained dielectric and NMR correlation times, the latter from $T_1$ and stimulated-echo measurements. They were compared with previously published results from differential scanning calorimetry and dielectric spectroscopy. A large deviation from an Arrhenius law was found, compatible with a steepness index of m = 79. Using dielectric measurements also the β-relaxation was studied and shown to exhibit the typical signatures of a JG relaxation. For $T < T_g$ the reorientation of the methyl group from $T_1(T)$ was found to be thermally activated. The differences in the measured activation energies in the glass with respect to those in the crystal can be understood by taking into account the broad energy barrier distribution in the glass. Thus, the widely used pharmaceutical ASA was found to be a model glass former that allows to study the heterogeneous dynamics over a particularly large temperature range.

We thank B. Geil for valuable suggestions and G.P. Johari for drawing our attention to the glass-forming ability of ASA, for sending us Ref. 7, and for useful comments, also on the manuscript. This work was supported by the Deutsche Forschungsgemeinschaft within the Graduiertenkolleg 298.

**Figure Captions:**

FIG. 1. Frequency dependent dielectric loss factor of ASA-$d_3$. The upturn seen at 230 K and low frequencies is due to the high-frequency flank of the α-relaxation. The lines for T ≥ 170 K are fits using the Cole-Cole expression and the parameters given in the text.

FIG. 2. Arrhenius plot of the correlation times of methyl deuterated ASA (closed symbols) and of fully protonated ASA (open symbols). They were obtained from dielectric measurements at 1 kHz (▽, Ref. 7) and in a broader range (▲ and ○, this work), from differential scanning calorimetry (DSC) at 100 K/min (□, Ref. 6) and at 10 K/min (◇, Ref. 5), from stimulated-echo NMR (●, Ref. 20) and from $T_1$ experiments (◆, this work). The dash-dotted line represents an Arrhenius law and the solid line a Vogel-Fulcher law with the parameters given in the text. The DSC results of Ref. 6 depended somewhat on the heating rate which was varied between 20 and 150 K/min. Here we show those data which approach the dielectric ones closest. All results described so far were obtained for supercooled or glassy ASA. The closed squares refer to $CD_3$ group reorientation times in crystalline ASA as reported in Ref. 11 with the dashed line representing an energy barrier of 510 K.

FIG. 3. (a) Spin-lattice relaxation times and (b) stretching exponents of ASA-$d_3$. The results on the supercooled and glassy samples (●,▲) are compared with the data for polycrystals (▽, this work and +, Ref. 12) and a single crystal (△, Ref. 11). The dashed line is drawn to guide the eye. The other lines reflect Arrhenius laws with barriers of 510 K (dotted lined), 410 K (dash-dotted line), and 270 K (solid line).

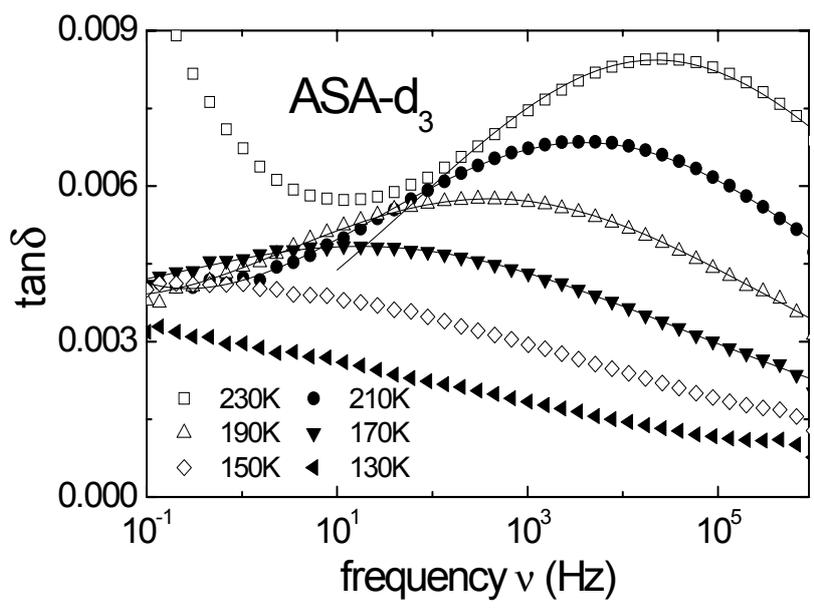

Fig. 1, R. Nath et al.

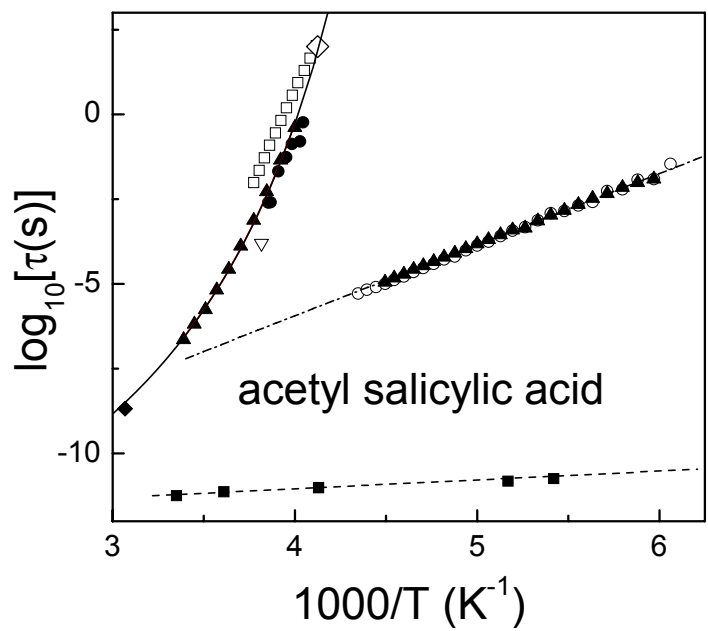

Fig. 2, R. Nath et al.

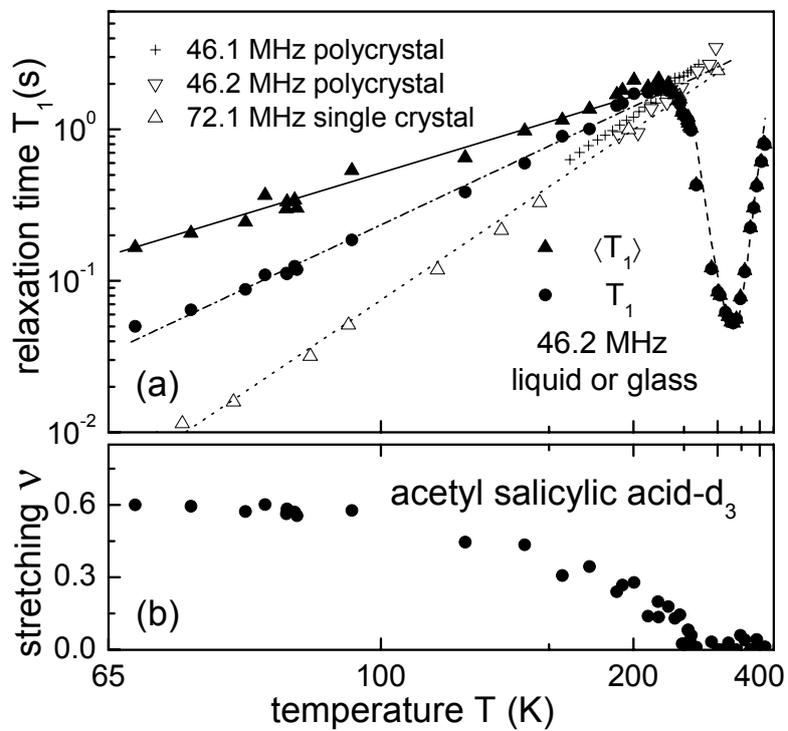

Fig. 3, R. Nath et al.